\documentstyle[12pt,epsfig]{article}

\newcommand{\be}{\begin{equation}}
\newcommand{\ee}{\end{equation}}
\newcommand{\Dlt}{\Delta}
\newcommand{\dlt}{\delta}

\newcommand{\br}{{\bf r}}
\newcommand{\bk}{{\bf k}}

\newcommand{\bP}{{\bf P}}
\newcommand{\bp}{{\bf p}}

\newcommand{\bt}{\beta}
\newcommand{\vp}{\varphi}
\newcommand{\ep}{\varepsilon}
\newcommand{\al}{\alpha}
\newcommand{\ra}{\rightarrow}
\newcommand{\sgm}{\sigma}

\newcommand{\gm}{\gamma}
\newcommand{\om}{\omega}
\newcommand{\Om}{\Omega}
\newcommand{\Gm}{\Gamma}
\newcommand{\dgr}{\dagger}
\newcommand{\lbd}{\lambda}

\begin{document}

\begin{center}
{\Large{\bf Bose systems in spatially random or time-varying 
potentials} \\ [5mm]

V.I. Yukalov$^1$, E.P. Yukalova$^2$, and V.S. Bagnato$^3$} \\ [3mm]

{\it $^1$Bogolubov Laboratory of Theoretical Physics, \\
Joint Institute for Nuclear Research, Dubna 141980, Russia\\ [2mm]

$^2$Laboratory of Information Technologies,\\
Joint Institute for Nuclear Research, Dubna 141980, Russia\\ [2mm]

$^3$Instituto de Fisica de S\~ao Calros, \\
Universidade de S\~ao Paulo, CP 369,
13560-970 S\~ao Carlos, S\~ao Paulo, Brazil\\
 }

\end{center}

\vskip 1cm

E-mail: yukalov@theor.jinr.ru

\vskip 2cm

\begin{abstract}

Bose systems, subject to the action of external random potentials,
are considered. For describing the system properties, under the action
of spatially random potentials of arbitrary strength, the stochastic
mean-field approximation is employed. When the strength of disorder
increases, the extended Bose-Einstein condensate fragments into
spatially disconnected regions, forming a granular condensate. Increasing
the strength of disorder even more transforms the granular condensate
into the normal glass. The influence of time-dependent external potentials
is also discussed. Fastly varying temporal potentials, to some extent,
imitate the action of spatially random potentials. In particular, strong
time-alternating potential can induce the appearance of a nonequilibrium
granular condensate.

\end{abstract}

\newpage

\section{Introduction}

Physics of systems with a Bose-Einstein condensate in random media
has been attracting attention for many years. At the beginning, the
interest was concentrated on the behavior of liquid helium in nanoporous
media [1]. In recent years, the physics of dilute Bose gases has gained
much interest [2--8]. Quasidisordered Bose gases have been realized
experimentally  by creating quasiperiodic optical lattices [9--11].
Disordered optical lattices have been realized by incorporating random
impurities into an optical lattice [12]. There exists vast literature
devoted to the theory of disordered optical lattices. A large number
of references on the subject can be found in the recent review article
[13].

A different type of disordered Bose systems corresponds to the systems
without a periodic lattice potential, but which are subject to a
spatially random potential. Experimentally, such random potentials
are realized by means of optical speckles [14--17]. Physical
properties of these Bose systems in an external random potential
have been theoretically studied for the case of weak disorder
[18--21] and for arbitrarily strong disorder [22,23].

The properties of uniform Bose systems and of those inside periodic
lattice potentials are, of course, different [13] and may require
differing theoretical methods of description. For example, the
self-consistent mean-field approximation [24--29] describes well
the uniform Bose systems with any strong atomic interactions. But
the region of applicability of this approximation can be limited
for Bose atoms in a lattice at zero temperature in the vicinity of
the superfluid-insulator phase transition [30].

The properties of disordered Bose systems are also different,
depending on whether external random potentials have been imposed
on an initially uniform or spatially periodic system [13]. One
interesting feature, common for both types of systems, is that
sufficiently strong disorder can lead to the occurrence of a new
phase, called Bose glass. This state of matter, occurring in
disordered lattices, was suggested by Fisher et al. [31] (see also
the recent articles [32,33] and the review paper [13]). The Bose 
glass phase is often characterized by the remaining presence of the
condensate fraction ($n_0>0$), but with the absence of superfluidity
($n_s=0$).

The nature of the Bose glass phase is usually described as follows.
The whole system fragments into the islands of the Bose-Einstein
condensate localized in the deep wells of the random potential, while
other regions, surrounding these islands, are filled by the normal
fluid, containing no condensate. This is why such a phase is also
termed the {\it granular condensate} or the {\it localized condensate}.
This phase has been observed in disordered optical lattices [34] and
in nanoporous media filled by liquid helium [35]. Such a state of
matter can also exist in Bose systems without lattices [13,36].

The present paper concerns the Bose systems without periodic lattices.
In the absence of a random potential, such a system would be uniform,
exhibiting at low temperatures Bose-Einstein condensation. The
description of a Bose-condensed system in external spatially random
potentials of arbitrary strength can be accomplished by means of the
stochastic mean-field approximation. We emphasize that describing such
random systems requires a special caution, since perturbation theory
is not always applicable to them. This, e.g., concerns the weak-disorder
expansion, which can become invalid for random Bose-condensed systems.
For these systems, the perturbation theory with respect to the coupling
parameter can also fail. Therefore the use of more refined approaches,
such as the stochastic mean-field approximation, is of great importance
for the correct description of random systems.

After presenting the main characteristics of Bose-condensed systems
in spatial random potentials, we turn to the problem of these systems
in {\it temporal} alternating potentials. We show that the action of
temporal external fields can induce in a Bose system the consequences
analogous to those produced by spatially random potentials. In particular,
a {\it nonequilibrium granular condensate} can arise under sufficiently
strong external fields. Very strong alternating fields will destroy the
granular condensate, transforming the whole system into a normal turbulent
fluid.

Throughout the paper, the natural system of units is employed, where
$\hbar=1$ and $k_B=1$.

\section{Bose Systems in Random Potentials}

We consider a dilute Bose system composed of atoms interacting through
the local potential
\be
\label{1}
\Phi(\br) = \Phi_0\dlt(\br) \; , \qquad
\Phi_0 \equiv 4\pi\; \frac{a_s}{m} \; ,
\ee
in which $a_s$ is the scattering length and $m$, atomic mass. The system
is subject to a spatially random external potential $\xi(\br)$. So that
the energy Hamiltonian is
\be
\label{2}
\hat H = \int \psi^\dgr(\br) \left [ - \; \frac{\nabla^2}{2m} +
\xi(\br) \right ] \; \psi(\br) \; d\br \; + \;
\frac{\Phi_0}{2} \; \int \psi^\dgr(\br) \psi^\dgr(\br)
\psi(\br) \psi(\br) \; d\br \; ,
\ee
where $\psi(\br)$ is the Bose field operator.

Without the loss of generality, the random potential can be taken as
zero-centered, such that its stochastic averaging gives
\be
\label{3}
\ll \xi(\br) \gg \; = \; 0 \; .
\ee
The stochastic correlation function
\be
\label{4}
R(\br - \br') \; = \; \ll \xi(\br) \xi(\br') \gg
\ee
is assumed to be real and symmetric,
\be
\label{5}
R^*(\br) = R(-\br) = R(\br) \; .
\ee

The presence of a Bose-Einstein condensate implies the spontaneous gauge
symmetry breaking [8], which is standardly realized through the Bogolubov
shift [37] of the field operator
\be
\label{6}
\psi(\br ) \; \ra \; \hat \psi(\br) \equiv
\eta(\br) + \psi_1(\br) \; .
\ee
Here $\eta(\br)$ is the condensate wave function normalized to the number
of condensed atoms,
\be
\label{7}
N_0 = \int |\eta(\br)|^2 \; d\br \; ,
\ee
and $\psi_1(\br)$ is the Bose field operator of uncondensed atoms defining
the number operator
\be
\label{8}
\hat N_1 = \int \psi_1^\dgr (\br) \psi_1(\br) \; d\br
\ee
of uncondensed atoms. The field variables $\eta(\br)$ and $\psi_1(\br)$
are treated as two independent variables, orthogonal to each other,
\be
\label{9}
\int \eta^*(\br) \psi_1(\br) \; d\br  = 0 \; .
\ee
Under the Bogolubov shift (6), in order to make the theory self-consistent,
it is necessary, as is proved in Refs. [24--29], to introduce the grand
Hamiltonian
\be
\label{10}
H = \hat H - \mu_0 N_0 - \mu_1 \hat N_1 \; ,
\ee
in which $\hat H$ is the energy Hamiltonian (2), with the
Bogolubov-shifted field operators (6), and $\mu_0$ and $\mu_1$ are the
Lagrange multipliers guaranteeing the theory self-consistency.

In the presence of random fields, there exist two types of averaging.
One type is the stochastic averaging, denoted as $\ll\ldots\gg$, which
characterizes the averaging over the distribution of random potentials.
And there is, as usual, the quantum statistical averaging, which for an
operator $\hat A$ is defined as
\be
\label{11}
< \hat A>_H \; \equiv \; {\rm Tr}\; \hat\rho \hat A \; ,
\ee
where $\hat\rho$ is a statistical operator. The latter, for an equilibrium
system, is
\be
\label{12}
\hat\rho = \frac{\exp(-\bt H)}{{\rm Tr}\exp(-\bt H)} \; ,
\ee
with $\bt\equiv 1/T$ being inverse temperature. The total average of an
operator $\hat A$,
\be
\label{13}
< \hat A> \; \equiv \; \ll {\rm Tr}\; \hat\rho \hat A \gg
\ee
includes both, the quantum and stochastic averaging procedures. The grand
thermodynamic potential, corresponding to the frozen disorder, is
\be
\label{14}
\Om = - T \ll \ln \; {\rm Tr}\; e^{-\bt H} \gg \; .
\ee

One should also keep in mind the quantum-number conservation condition
\be
\label{15}
< \psi_1(\br) > \; = \; 0 \; .
\ee
To satisfy the latter, the grand Hamiltonian (10) should be complimented
by one more term guaranteeing the absence in $H$ of the terms linear in
$\psi_1(\br)$, as is shown in Ref. [29]. Here we do not add explicitly such a
linear killer, since for a uniform system or for a system uniform on average,
linear in $\psi_1(\br)$ terms do not arise and condition (15) is
automatically satisfied [24--29].

For the zero-centered random potential, for which property (3) holds, the
system can be treated as uniform on average. Then one can set
\be
\label{16}
\eta(\br) = \sqrt{\rho_0} \qquad
\left ( \rho_0 \equiv \frac{N_0}{V} \right ) \; ,
\ee
where $V$ is the system volume. The field operator of uncondensed atoms 
can be expanded over plane waves as
\be
\label{17}
\psi_1(\br) = \frac{1}{\sqrt{V} } \;
\sum_{k\neq 0} a_k e^{i\bk\cdot\br} \; .
\ee
The field operators in the momentum representation, $a_k$, define the momentum
distribution
\be
\label{18}
n_k \; \equiv \; < a_k^\dgr a_k >
\ee
and the anomalous average
\be
\label{19}
\sgm_k \; \equiv \; < a_k a_{-k} > \;.
\ee
The density of uncondensed atoms is
\be
\label{20}
\rho_1  = \frac{1}{V}\; \sum_{k\neq 0} n_k \; ,
\ee
and the total anomalous average is
\be
\label{21}
\sgm_1 = \frac{1}{V} \; \sum_{k\neq 0} \sgm_k \; .
\ee
With the total number of particles $N$, the system average density is
\be
\label{22}
\rho \equiv \frac{N}{V} = \rho_0 + \rho_1 \; .
\ee
The density of uncondensed atoms (20) consists of two terms,
\be
\label{23}
\rho_1 = \rho_N + \rho_G \; ,
\ee
among which $\rho_N$ is due to thermal fluctuations and interactions, 
while $\rho_G$ is caused by the random potential and defines the density 
of a glassy component or, briefly speaking, the {\it glassy density}
\be
\label{24}
\rho_G \equiv \frac{1}{V} \; 
\int \ll | < \psi_1(\br)>_H  |^2 \gg \; d\br \; .
\ee
The general definition of the superfluid density [3,29] results in the 
expression
\be
\label{25}
\rho_s = \rho \; - \; \frac{<\hat\bP^2>}{3mTV} \; ,
\ee
in which $\hat\bP$ is the momentum operator
$$
\hat\bP \equiv \int \hat\psi(\br) \; (-i\nabla ) \;
\hat \psi(\br) \; d\br = \int \psi_1(\br) \; ( - i\nabla) \;
\psi_1(\br) \; d\br \; .
$$

It is convenient to define the dimensionless atomic fractions. Thus, the
{\it condensate fraction}
\be
\label{26}
n_0 \equiv \frac{\rho_0}{\rho} = 1 - n_1
\ee
is expressed through the {\it uncondensed-atom fraction}
\be
\label{27}
n_1 \equiv \frac{\rho_1}{\rho} = \frac{1}{N} \; \sum_{k\neq 0} n_k \; .
\ee
Equation (21) gives the {\it anomalous fraction}
\be
\label{28}
\sgm \equiv \frac{\sgm_1}{\rho} = \frac{1}{N} \;
\sum_{k\neq 0} \sgm_k \; .
\ee
The uncondensed-atom fraction (27) is the sum
\be
\label{29}
n_1 = n_N + n_G
\ee
of the {\it normal fraction} $n_N\equiv\rho_N/\rho$ and the {\it glassy
fraction}
\be
\label{30}
n_G \equiv \frac{\rho_G}{\rho} = \frac{1}{N} \;
\int \ll | < \psi_1(\br)>_H |^2 \gg \; d\br \; .
\ee
Finally, from Eq. (25) we have the {\it superfluid fraction}
\be
\label{31}
n_s = 1 \; - \; \frac{2Q}{3T} \; ,
\ee
in which
\be
\label{32}
Q \equiv \frac{<\hat\bP^2> }{2mN}
\ee
is the dissipated heat per atom. With expansion (17), the glassy fraction 
can be written as
\be
\label{33}
n_G = \frac{1}{N} \; \sum_k \ll |\al_k|^2 \gg \; ,
\ee
where
\be
\label{34}
\al_k \; \equiv < a_k>_H \; .
\ee

Together with expanding over plane waves the field operator of uncondensed 
atoms, as in Eq. (17), let us expand the random potential as
\be
\label{35}
\xi(\br) =\frac{1}{\sqrt{V} } \; \sum_k \xi_k e^{i\bk\cdot\br} \; , 
\qquad \xi_k = \frac{1}{\sqrt{V}} \; 
\int \xi(\br) e^{-i\bk\cdot\br} \; d\br \; .
\ee
The Fourier transformation of the correlation function (4) is
\be
\label{36}
R(\br) = \frac{1}{V} \; \sum_k R_k e^{i\bk\cdot\br} \; , \qquad
R_k = \int R(\br) e^{-i\bk\cdot\br} \; d\br \; .
\ee
Then, Fourier-transforming Eq. (4), we get
\be
\label{37}
\ll \xi_k^* \xi_p\gg \; = \; \dlt_{kp} R_k \; .
\ee

Passing to the momentum representation, we accomplish in Hamiltonian (2) 
the Bogolubov shift (6) and substitute there the Fourier transforms (17) 
and (35). As a result, the grand Hamiltonian (10) acquires the form
\be
\label{38}
H = \sum_{n=0}^4 H^{(n)} \; + \; H_{ext} \; ,
\ee
in which
$$
H^{(0)} = \left ( \frac{1}{2} \; \rho_0 \Phi_0 - \mu_0 \right ) N_0 \; ,
\qquad H^{(1)} = 0 \; ,
$$
$$
H^{(2)} = \sum_{k\neq 0} \left ( \frac{k^2}{2m} + 2\rho_0 \Phi_0 - 
\mu_1\right ) a_k^\dgr a_k \; + \; \frac{1}{2} \; \sum_{k\neq 0} 
\rho_0 \Phi_0 \left ( a_k^\dgr a_{-k}^\dgr + a_{-k} a_k \right ) \; ,
$$
$$
H^{(3)} = \; \sqrt{\frac{\rho_0}{V} } \; 
\sum_{k,p(\neq 0)} \; \Phi_0 \left ( a_k^\dgr a_{k+p} a_{-p} + 
a_{-p}^\dgr a_{k+p}^\dgr a_k \right ) \; ,
$$
\be
\label{39}
H^{(4)} = \frac{1}{2V} \; \sum_q \; \sum_{k,p(\neq 0)} \; 
\Phi_0 a_k^\dgr a_p^\dgr a_{k-q} a_{p+q} \; ,
\ee
with the part
\be
\label{40}
H_{ext} = \rho_0 \xi_0 \sqrt{V} \; +\; \sqrt{\rho_0} \; \sum_{k\neq 0}
\left ( a_k^\dgr \xi_k + \xi_k^* a_k \right ) \; + \;
\frac{1}{\sqrt{V}} \; \sum_{k,p(\neq 0)} \; a_k^\dgr a_p \xi_{k-p} \; ,
\ee
which is due to the action of the external random potential.

\section{Stochastic Mean-Field Approximation}

To treat Hamiltonian (38), an approximation is needed. The third- and
fourth-order terms in the operators $a_k$, occurring in Eq. (39), can be 
simplified by means of the Hartree-Fock-Bogolubov approximation, as in 
Refs. [24--29]. However, the interaction of the random potential with 
atoms, described by part (40), cannot be treated in the simple mean-field 
procedure, since
\be
\label{41}
< a_k > \; = \; < \xi_k > \; = \; 0 \; ,
\ee
which would kill the last term in Eq. (40). Such a case would correspond 
to considering asymptotically weak disorder. To treat the last term in 
(40), a more delicate decoupling procedure is required. For this purpose, 
we shall employ the {\it stochastic mean-field approximation} suggested 
and used earlier for other physical systems [38--43]. Following the idea 
of this approximation, we simplify the last term of Eq. (40) writing
\be
\label{42}
a_k^\dgr a_p \xi_{k-p} = \left ( a_k^\dgr \al_p + 
\al_k^* a_p - \al_k^* \al_p\right ) \xi_{k-p} \; ,
\ee
where $\al_k$ is the quantum average (34). Under equation (42), we have
\be
\label{43}
< a_k^\dgr a_p \xi_{k-p} > \; = \; \ll 
\al_k^* \al_p \xi_{k-p} \gg \; .
\ee
Approximations (42) and (43) allow for the consideration of arbitrarily 
strong disorder strengths [22,23].

The next nontrivial procedure is the diagonalization of Hamiltonian (38) 
by means of the {\it nonuniform nonlinear} canonical transformation
\be
\label{44}
a_k = u_k b_k + v_{-k}^* b_{-k}^\dgr + w_k \vp_k \; ,
\ee
in which $u_k$, $v_k$, $w_k$, and $\vp_k$ are to be defined so that the 
resulting Hamiltonian be diagonal in the operators $b_k$, such that
\be
\label{45}
< b_k >_H \; = \; < b_k b_p>_H \; = \; 0 \; .
\ee
Then the transformation (44) gives
\be
\label{46}
\al_k \; \equiv \; < a_k >_H \; = \; w_k \vp_k \; .
\ee
Accomplishing the diagonalization, we find
\be
\label{47}
u_k^2 = \frac{\om_k+\ep_k}{2\ep_k} \; , \qquad 
v_k^2 = \frac{\om_k-\ep_k}{2\ep_k} \; , \qquad
u_kv_k = -\; \frac{mc^2}{2\ep_k} \; , \qquad
w_k = -\; \frac{1}{\om_k+mc^2} \; ,
\ee
where 
\be
\label{48}
\om_k \equiv \frac{k^2}{2m} + mc^2 \; , \qquad 
\om_k^2  = \ep_k^2 + \left ( mc^2\right )^2 \; ,
\ee
and $\ep_k$ is the Bogolubov spectrum
\be
\label{49}
\ep_k =\sqrt{(ck)^2 + \left ( 
\frac{k^2}{2m}\right )^2 } \; ,
\ee
with the sound velocity given by the equation
\be
\label{50}
mc^2  = (n_0 +\sgm) \rho \Phi_0 \; .
\ee
The random variable $\vp_k$ satisfies the integral equation
\be
\label{51}
\vp_k = \sqrt{\rho_0} \; \xi_k \; - \; \frac{1}{\sqrt{V}} \;
\sum_p \; \frac{\xi_{k-p}\vp_p}{\om_p + mc^2 }
\ee
of the Fredholm type. The nonuniform nonlinear transformation (44), 
with Eqs. (47) to (51), results in the grand Hamiltonian
\be
\label{52}
H = E_B  \; + \; \sum_k \ep_k b_k^\dgr b_k \; + \; 
\vp_0 \; \sqrt{N_0} \; ,
\ee
in which
$$
E_B = \frac{1}{2} \; \sum_k ( \ep_k - \om_k) \; - \; 
\left [ 1 - n_0 ( 1 +\sgm) + \frac{1}{2}\left ( 1 -n_1^2 + 
\sgm^2 \right ) \right ] \; \rho^2 \Phi_0 N \; .
$$

With the diagonal Hamiltonian (52), it is straightforward to find 
the momentum distribution (18),
\be
\label{53}
n_k = \frac{\om_k}{2\ep_k} \; {\rm coth} \left (
\frac{\ep}{2T} \right ) \; - \; \frac{1}{2} \; + \;
\ll |\al_k|^2 \gg
\ee
and the anomalous average
\be
\label{54}
\sgm_k = -\; \frac{mc^2}{2\ep_k} \; {\rm coth} \left (
\frac{\ep_k}{2T} \right ) \; + \; \ll |\al_k|^2 \gg \; .
\ee
The random variable (46), in view of Eqs. (47), is
\be
\label{55}
\al_k = - \; \frac{\vp_k}{\om_k+mc^2} \; .
\ee
>From here
\be
\label{56}
\ll |\al_k|^2 \gg \; = \; 
\frac{\ll|\vp_k|^2\gg}{(\om_k+mc^2)^2} \; .
\ee

The fraction of uncondensed atoms (27) is represented as sum (29), 
in which
\be
\label{57}
n_N = \frac{1}{2\rho} \; \int \left [ \frac{\om_k}{\ep_k} \;
{\rm coth} \left ( \frac{\ep_k}{2T} \right ) - 1 \right ] \; 
\frac{d\bk}{(2\pi)^3} \; ,
\ee
while the glassy fraction (33) becomes
\be
\label{58}
n_G = \frac{1}{\rho} \; \int 
\frac{\ll |\vp_k|^2\gg}{(\om_k+mc^2)^2} \; \frac{d\bk}{(2\pi)^3} \; .
\ee
The anomalous fraction (28) is also the sum
\be
\label{59}
\sgm = \sgm_N + n_G \; ,
\ee
where
\be
\label{60}
\sgm_N = -\; \frac{1}{2\rho} \; \int \; \frac{mc^2}{\ep_k} \;
{\rm coth}\left ( \frac{\ep_k}{2T} \right ) \; \frac{d\bk}{(2\pi)^3} 
\ee
and the $n_G$ is the same as in Eq. (58). The terms $n_N$ and $\sgm_N$ 
are caused by thermal fluctuations and interactions, while the glassy 
fraction $n_G$ is due to the action of the random field.

In the superfluid fraction (31), the dissipated heat (32) reads as
\be
\label{61}
Q = Q_N + Q_G \; ,
\ee
with
\be
\label{62}
Q_N = \frac{1}{8m\rho} \; \int \;
\frac{k^2}{{\rm sinh}^2(\ep_k/2T)} \;
\frac{d\bk}{(2\pi)^3} 
\ee
being due to interactions and finite temperature, while
\be
\label{63}
Q_G = \frac{1}{2m\rho} \; \int \;
\frac{k^2\ll|\vp_k|^2\gg}{\ep_k(\om_k+mc^2)} \;
{\rm coth} \left ( \frac{\ep_k}{2T} \right ) \; \frac{d\bk}{(2\pi)^3} 
\ee
is the heat dissipated by the glassy fraction.

Note that deriving the above formulas we have used the form of the 
Lagrange multiplier
\be
\label{64}
\mu_0 = (1 + n_1 +\sgm) \rho \Phi_0 \; ,
\ee
obtained by minimizing the grand potential (14) with respect to the 
number of condensed atoms $N_0$, and the multiplier
\be
\label{65}
\mu_1 =  ( 1 + n_1 - \sgm) \rho \Phi_0 \; ,
\ee
found from the condition of the condensate existence [13], which 
is equivalent to the requirement that the spectrum of collective 
excitations be gapless.

\section{Failure of Weak-Disorder Perturbation Theory}

Uniform Bose-condensed systems under the action of random 
potentials are usually considered in the case of asymptotically 
weak interactions and weak disorder. Exceptions are the articles 
[22,23], where the stochastic mean-field approximation [38--43] 
was employed allowing for the description of systems with arbitrarily 
strong atomic interactions and arbitrarily strong disorder. It is 
necessary to stress that the perturbation theory with respect to 
the disorder strength may be inapplicable to the Bose systems in 
random potentials. In the present section, we show that the 
weak-disorder perturbation expansion may lead to incorrect results.

Let us consider the last term
\be
\label{66}
H_{ran} \equiv \vp_0 \; \sqrt{N_0} 
\ee
in Hamiltonian (52), explicitly describing the influence of the 
random potential. The related contribution to the internal energy 
is
\be
\label{67}
E_{ran} \; \equiv \; < H_{ran} >\; ,
\ee
which results in
\be
\label{68}
E_{ran} \; =\; < \vp_0>\sqrt{N_0} \; = \; 
\ll \vp_0\gg \sqrt{N_0} \; .
\ee
Suppose, one considers weak disorder and assumes that the 
weak-disorder perturbation theory should be valid. Then, the first 
term in Eq. (51) can be treated as the zero-order approximation 
\be
\label{69}
\vp_k^{(0)} = \sqrt{\rho_0}\; \xi_k \; .
\ee
Iterating with Eq. (69) the second term of Eq. (51) gives the 
first-order approximation
\be
\label{70}
\vp_k^{(1)} = \sqrt{\rho_0}\; \xi_k \; -\; 
\frac{\sqrt{N_0}}{V} \;
\sum_p \; \frac{\xi_{k-p}\xi_p}{\om_p+mc^2} \; .
\ee
Employing approximation (70) makes it easy to derive all results 
obtained by other researchers using the weak-disorder perturbation 
theory (see Ref. [22]). For example, the internal energy (68), due 
to random fields, becomes
\be
\label{71}
E_{ran}^{(1)} = - \; \frac{N_0}{V} \; \sum_p \; 
\frac{\ll |\xi_p|^2| \gg}{\om_p+mc^2} \; .
\ee
Using here the correlation formula (37) and passing to integration 
yields
\be
\label{72}
E_{ran}^{(1)} = - \int \; \frac{N_0R_p}{(\om_p+mc^2)} \; 
\frac{d\bp}{(2\pi)^3} \; .
\ee
It is exactly this expression (72) that has been rederived by 
all authors who have used the weak-disorder perturbation theory. 
Equation (72) tells us that random fields diminish the internal 
energy.

But the result is very different, if no perturbation theory has 
been involved. According to Eqs. (34) and (41), we have
\be
\label{73}
< a_k> \; = \; \ll \al_k \gg \; = \; 0 \; .
\ee
Then, from Eq. (46), it follows that
\be
\label{74}
\ll \vp_k \gg \; = \; 0 \; .
\ee
Consequently, the random-field contribution to the internal 
energy (68) is {\it exactly} zero,
\be
\label{75}
E_{ran} \; \equiv \; < H_{ran}>\; = \; 0\; .
\ee
This conclusion sounds reasonable, if one remembers that the 
zero-centered potential, with the zero mean (3), is considered. 
Thus, the perturbation-theory formula (72) is in contradiction 
with the exact expression (75), hence, Eq. (72) does not seem 
to be correct. The same concerns other formulas that have been 
derived by means of the weak-disorder expansion. Such formulas 
do not seem to be reliable. The reason for the weak-disorder 
perturbation theory failure is the noncommutativity, in some 
cases, of the operations of expanding over the random fields 
and of averaging over these fields.

\section{Random Potential with Local Correlations}

The consideration of the previous sections is general, beeing 
applicable to any spatially random potentials. To specify the 
problem, it is necessary to concretize the type of the random 
potential. When the correlation length of the correlation 
function (4) is sufficiently short, at least much shorter than 
the healing length, then the random potential can be modelled 
by the Gaussian white noise, with the local correlation function
\be
\label{76}
R(\br) = R_0 \dlt(\br) \; .
\ee
Then the Fourier transform (36) gives $R_k=R_0$ and Eq. (37) 
becomes
\be
\label{77}
\ll \xi_k^* \xi_p\gg \; =\; \dlt_{kp} R_0 \; .
\ee

For numerical investigation, it is convenient to introduce 
dimensionless variables. The interaction strength is characterized 
by the {\it gas parameter}
\be
\label{78}
\gm \equiv \rho^{1/3} a_s \; .
\ee
The dimensionless temperature is defined as
\be
\label{79}
t \equiv \frac{mT}{\rho^{2/3}}\; .
\ee
For the dimensionless sound velocity, we have
\be
\label{80}
s \equiv \frac{mc}{\rho^{1/3} }\; .
\ee

Disorder is known to possess the property of localizing atomic 
motion inside the regions of the characteristic {\it localization 
length}, which can be estimated as the Larkin length [44]
\be
\label{81}
l_{loc} \equiv \frac{4\pi}{7m^2 R_0} \; .
\ee
The strength of disorder can be described by the {\it disorder 
parameter}
\be
\label{82}
\zeta \equiv \frac{a}{l_{loc}} \qquad 
\left ( \rho^{1/3} a =  1 \right ) \; ,
\ee
where $a$ is the mean interatomic distance.

If the random potential $\xi(\br)$ is limited by a finite amplitude 
$V_R$, such that
$$
|\xi(\br)| \leq V_R
$$
for all $\br$ pertaining to the considered system, and if the 
correlation function (4) has a finite correlation length $l_R$, 
then the parameter $R_0$ in Eqs. (76) and (77) can be represented 
as
\be
\label{83}
R_0 =  V_R^2\; l_R^3 \; .
\ee
In that case, the localization length (81) becomes
\be
\label{84}
l_{loc} = \frac{4\pi}{7m^2 V_R^2\; l_R^3} \; ,
\ee
which is close to the expression given in Ref. [17]. Returning 
to the local correlation function (76) implies that $l_R\ra 0$ 
and $V_R\ra\infty$, so that product (83) be finite.

When disorder is asymptotically weak, so that $R_0\ra 0$, then 
$l_{loc}$ extends to the size of the whole system. The latter 
then represents the standard Bose-condensed system with extended 
condensate. With growing disorder, when the localization length 
becomes much smaller than the system linear size $L$, but when 
$l_{loc}$ is yet much larger than the mean interatomic distance 
$a$, that is, when
\be
\label{85}
a \ll l_{loc} \ll L \; ,
\ee
then the Bose-Einstein condensate fragments into multiple 
pieces separated by the normal phase with no gauge symmetry 
breaking [34--36,45--47]. This type of matter is termed the 
{\it Bose glass} or the {\it granular condensate}. Finally, 
when disorder is so strong that $l_{loc}\sim a$, then no 
condensate is possible, since atoms are localized separately, 
each of them being trapped in a deep randomly located well of 
the random potential. The latter phase forms the normal glassy 
matter. Such a phase was observed [14] in a strong random 
potential created by laser speckles. At zero temperature, the 
phase portrait, in the variables of the gas parameter (78) and 
disorder parameter (82), should look as in Fig. 1. The granular 
condensate starts appearing when $l_{loc}\sim(10-100)a$, hence 
$\zeta\sim 0.01-0.1$. And the granular condensate transforms 
into the normal glass when $l_{loc}\sim a$, so that $\zeta\sim 
1$. The phase transition between the extended condensate and 
granular condensate is continuous, while that between the 
granular condensate and the normal glass is of first order.

The granular condensate is a phase that cannot already be 
treated as uniform on average, as it is done for the extended 
condensate. The granular condensate is a principally nonuniform 
condensate, which requires a separate consideration taking into 
account the spatial nonuniformity. Assuming that the system is 
uniform on average does not distinguish between the extended 
and granular condensates, but this description covers all the 
region, where any condensate is possible. However, the phase 
transition between the system with a condensate and the normal 
glass can be described. Keeping this in mind, we follow the 
consideration of the previous sections.

First, we need to solve the integral equation (51). A good 
approximate solution to this equation  is
\be
\label{86}
\vp_k = \frac{\sqrt{\rho_0}\; \xi_k}{1 + 
\frac{1}{\sqrt{V}}\sum_p\frac{\xi_p}{\om_p+mc^2} } \; .
\ee
This random variable enters the above equations, such as Eqs. 
(53), (54), (55), (58), and (63), in the form $\ll|\vp_k|^2\gg$. 
To find the latter, we employ the self-similar approximation 
theory [48--50] in the variant involving the self-similar factor 
approximants [51--54]. As a result, we obtain
\be
\label{87}
\ll |\vp_k|^2 \gg \; = \; 
\frac{n_0 R_0 s^{3/7}}{a^3(s-\zeta)^{3/7}} \; .
\ee
Then the glassy fraction (58) is
\be
\label{88}
n_G = \frac{n_0\zeta}{7s^{4/7}(s-\zeta)^{3/7} } \; .
\ee
Note that the disorder parameter (82) naturally appears in Eqs. 
(87) and (88). Therefore these equations can be considered as 
defining the disorder parameter (82) as such.

In dimensionless variables (78) to (80), the dimensional equation 
(50) for the sound velocity becomes
\be
\label{89}
s^2 =  4\pi\gm (1 - n_1 +\sgm) \; .
\ee
Due to Eqs. (29) and (59), this can be rewritten as
\be
\label{90}
s^2 =  4\pi\gm (1 - n_N + \sgm_N) \; .
\ee
For the normal fraction (57), we have
\be
\label{91}
n_N =\frac{s^3}{3\pi^2} \left \{ 1 +
\frac{3}{2\sqrt{2}} \; \int _0^\infty \; \left ( 
\sqrt{1+x^2} - 1 \right )^{1/2} \left [ 
{\rm coth}\left ( \frac{s^2 x}{2t}\right ) -1 
\right ] \; dx \right \} \; .
\ee
According to expressions (22) and (23), the atomic fractions are 
normalized as
\be
\label{92}
n_0 + n_N + n_G = 1 \; .
\ee
Using this, the glassy fraction (88) can be represented as
\be
\label{93}
n_G = \frac{(1-n_N)\zeta}{\zeta+7s^{4/7}(s-\zeta)^{3/7} } \; .
\ee

The superfluid fraction (31) contains the dissipated heat (61). 
The part (63) of the dissipated heat, due to the dissipation on 
the glassy fraction, in the case of the white noise, contains Eq. 
(87). Then, integral (63) diverges, but can be regularized [22]. 
So that, finally, for the superfluid fraction, we find
\be
\label{94}
n_s =  1 \; - \; \frac{4}{3}\; n_G \; - \; 
\frac{s^5}{6\sqrt{2}\;\pi^2 t} \; \int_0^\infty \;
\frac{x\left ( \sqrt{1+x^2}-1\right )^{3/2} \; dx}
{\sqrt{1+x^2}\;{\rm sinh}^2(s^2x/2t)}  \; .
\ee

The anomalous fraction (60), because of the local interaction 
(1), also diverges and requires to be regularized [13,24--29,55]. 
In this regularization, the asymptotic properties of function (60), 
with respect to temperature, should remain correctly defined. Thus, 
in the limit of zero temperature, using the dimensional 
regularization, we have [13,24,27--29]
\be
\label{95}
\sgm_N \simeq \frac{2s^2}{\pi^2}\; \sqrt{\pi\gm n_0}  
\qquad (t\ra 0) \; .
\ee
While, when temperature tends to the critical point
\be
\label{96}
t_c = 3.312498 \; ,
\ee
then the correct limit of Eq. (60) is
\be
\label{97}
\sgm_N \simeq -\; \frac{st}{2\pi} \qquad (t\ra t_c) \; .
\ee
The critical point (96) is the same as for the ideal Bose gas, 
as it has to be for a mean-field approximation [29].

It is important to stress that taking account of the anomalous 
averages is principally necessary. It would be absolutely 
incorrect to use the Shohno trick [56] by omitting the anomalous 
averages, as one often does. This is mathematically wrong, since 
the anomalous averages at low temperature can be larger than the 
normal fraction $n_N$, and they are of the same order at finite 
temperatures below the critical point [23,57]. Keeping what 
is of the same order, but neglecting what can even be larger, 
cannot be called a reasonable approximation. It is also 
straightforward to show [24,29] that omitting the anomalous 
average renders the consideration not self-consistent and the 
system unstable. The origin of the resulting inconsistency is 
very easy to understand [29]. The existence of the anomalous 
average is due to the gauge symmetry breaking. The latter is 
also the cause of the Bose-Einstein condensate existence. Hence 
both, the anomalous average and the condensate, either exist 
together or do not arise at all. If one wishes to omit the 
anomalous average, then, to be self-consistent, one must neglect 
the condensate existence. Or, when the latter is assumed, one has 
to retain the anomalous average as well. The omission of the 
anomalous average, in addition to breaking the system stability 
and making the thermodynamics not self-consistent, also distorts 
the phase transition order, provoking a first-order transition. 
The latter is, of course, incorrect, since the Bose-Einstein 
condensation is the {\it second-order} phase transiiton, 
irrespectively to the interaction strength [29].

The correct asymptotic behavior of the anomalous average 
in the vicinity of the critical point $t_c$, as in Eq. (97), 
guarantees the second-order phase transition for any value 
of the gas parameter [13,28,29]. This can be demonstrated by 
direct numerical calculations [13,28] and also by expanding 
the quantities of interest in powers of the relative 
temperature deviation
\be
\label{98}
\tau \equiv \left | 
\frac{t-t_c}{t_c} \right | \; \ra \; 0 
\ee
in the vicinity of $t_c$. Then we obtain the dimensionless 
sound velocity
\be
\label{99}
s \simeq \frac{3\pi}{t_c} \; \tau + \frac{9\pi}{t_c}
\left ( 1 \; - \; \frac{2\pi}{\gm t_c^2}
\right ) \; \tau^2 \; ,
\ee
the condensate fraction
\be
\label{100}
n_0 \simeq \frac{3}{2}\; \tau \; -\; 
\frac{3}{8} \; \tau^2 \; ,
\ee
the anomalous average
\be
\label{101}
\sgm_N \simeq -\; \frac{3}{2}\; \tau + \frac{3}{8} 
\left ( 1 + \frac{6\pi}{\gm t_c^2} \right ) \; \tau^2 \; ,
\ee
and the superfluid fraction
\be
\label{102}
n_s \simeq \frac{3}{2}\; \tau - 1.741 \; \tau^2 \; .
\ee
>From these asymptotic expressions, the second-order phase 
transition is evident.

Trying to interpolate the anomalous average between the 
asymptotic limits (95) and (97), we can reorganize Eq. (60) 
to the identical form
\be
\label{103}
\sgm_N = - \;\frac{1}{2\rho} \; \int \; 
\frac{mc^2}{\ep_k} \; \frac{d\bk}{(2\pi)^3} \;  - \; 
\frac{1}{2\rho} \; \int \; \frac{mc^2}{\ep_k} \left [ 
{\rm coth}\left ( \frac{\ep_k}{2T}\right ) - 1 
\right ] \; \frac{d\bk}{(2\pi)^3} \; .
\ee
Employing for the first term of Eq. (103) the dimensional
regularization and passing to the dimensionless quantities 
(78) to (80), we get [24,27--29]
\be
\label{104}
\sgm_N = \frac{2s^2}{\pi^2}\; \sqrt{\pi\gm n_0} \; - \; 
\frac{\sqrt{2}\;s^3}{(2\pi)^2} \;  \int_0^\infty \;
\frac{\left(\sqrt{1+x^2}-1\right )^{1/2}}{\sqrt{1+x^2}} \; 
\left [ {\rm coth}\left ( \frac{s^2x}{2t}\right ) -1
\right ] \; dx \; .
\ee
This expression gives the correct low-temperature limit (95). 
However, the critical behavior (97) becomes disturbed. This 
disturbance, as numerical calculations [28] show, is not 
essential for the gas parameter in the range $0<\gm\leq 0.3$, 
for which the phase transition remains continuous. For larger 
$\gm$, the use of expression (104) would result in the 
discontinuity at the critical point $t_c$. Therefore, for 
$\gm>0.3$, form (104) can be used below the critical region, 
while close to $T_c$, it has to be replaced by the correct 
limit (97) guaranteeing the continuous phase transition [28].

At zero temperature, the normal fraction (91) and the superfluid 
fraction (94) reduce to
\be
\label{105}
n_N = \frac{s^3}{3\pi^2} \; , \qquad 
n_s =  1 \; - \; \frac{4}{3}\; n_G \; .
\ee
These equations, together with Eqs. (90), (92), (93), and 
(95), have been analysed in detail [22,23], demonstrating a 
first-order phase transition on the line $\zeta=\zeta(\gm)$ 
between the system with a condensate and the normal glass with 
no condensate.

\section{Simple Model with Quenched Disorder}

In Sec. 4, we showed that the weak-disorder perturbation theory 
fails in the description of the Bose-condensed system in 
spatially random potentials. Therefore, for correctly describing 
such systems, more refined approaches are necessary, for instance, 
as the stochastic mean-field approximation of Sec. 3, which allows 
for the consideration of arbitrarily strong disorder.

Here we show that, in the presence of disorder, perturbation 
theory with respect to atomic interactions can also fail. We 
demonstrate this by a simple model with quenched disorder, 
which allows for an explicit illustration of the point where 
the weak-coupling perturbation theory fails.

Let us consider the model Hamiltonian
\be
\label{106}
H = ( 1+\xi) \vp^2 + g\vp^4 
\ee
of the so-called zero-dimensional $\vp^4$-theory in the presence 
of quenched disorder. Here the variable $\vp\in(-\infty,\infty)$ 
imitates the Bose condensate function. The coupling parameter $g>0$ 
characterizes the interaction strength. And the random variable 
$\xi$ describes an external random potential, with a distribution 
$p(\xi)$. As is usual, we assume that the random potential is 
zero-centered, such that
\be
\label{107}
\ll \xi \gg \; = \; 
\int_{-\infty}^{\infty} \; \xi p(\xi) \; d\xi = 0 \; .
\ee
Its dispersion is given by
\be
\label{108}
\Dlt^2 \; \equiv \; \ll \xi^2 \gg \; = \; 
\int_{-\infty}^{\infty} \xi^2 p(\xi) \; d\xi \; .
\ee

The often used examples of the random-variable distribution are 
the Gaussian distribution
\be
\label{109}
p_G(\xi) = \frac{1}{\sqrt{2\pi}\;\xi_0} \; \exp\left ( -\;
\frac{\xi^2}{2\xi_0^2} \right ) 
\ee
and the uniform distribution
\be
\label{110}
p_U(\xi) = \frac{1}{2D} \; \Theta(D-|\xi|) \; .
\ee
Their dispersions, respectively, are
$$
\Dlt_G^2 =\xi_0^2 \; , \qquad \Dlt_U^2 =\frac{D^2}{3} \; .
$$

The partition function of the system in the frozen random field is
\be
\label{111}
Z(g,\xi) = \frac{1}{\sqrt{\pi}} \; \int_{-\infty}^{\infty} \;
e^{-H} \; d\vp \; .
\ee
The free energy of the system with quenched disorder writes as
\be
\label{112}
f(g) = - \int_{-\infty}^{\infty} \; p(\xi) 
\ln Z(g,\xi)\; d\xi \; .
\ee

In the limit of no disorder, when $\Dlt^2\ra 0$, then, in both 
cases (109) and (110), one has $p(\xi)\ra\dlt(\xi)$. As a result,
$$
f(g) \ra -\ln Z(g,0) \qquad (\Dlt^2 \ra 0)
$$
with
$$
Z(g,0) = \frac{1}{\sqrt{\pi}} \; \int_{-\infty}^{\infty} \;
\exp\left (-\vp^2 - g\vp^4\right ) \; d\vp \; .
$$
The case of no disorder allows for the use of the weak-coupling 
perturbation theory, provided the latter is complimented by a 
resummation procedure [58,59].

However, for finite disorder, the situation is more complicated. 
The partition function (111), with Hamiltonian (106), reads as
\be
\label{113}
Z(g,\xi) = \frac{1}{\sqrt{\pi}} \; \int_{-\infty}^{\infty} \;
\exp\left \{ -(1+\xi) \vp^2 - g\vp^4\right \} \; d\vp \; .
\ee
Expanding here the integrand in powers of $g$ and integrating 
gives the series
\be
\label{114}
Z(g,\xi)  = \sum_{n=0}^\infty z_n(\xi) g^n
\ee
with the coefficients
\be
\label{115}
z_n(\xi) = 
\frac{(-1)^n\Gm(2n+1/2)}{\sqrt{\pi}\;
\Gm(n+1)(1+\xi)^{2n+1/2}} \; .
\ee
Substituting series (114) into the free energy (112) shows that 
the latter diverges for all $n=0,1,2,\ldots$, in the case of the 
Gaussian distribution , because of the pole at $\xi=-1$ in 
coefficient (115). In the case of the uniform distribution (110), 
integral (112) is finite only for weak disorder, for which $D<1$ 
and $\Dlt^2_U<1/3$. But for any stronger disorder, with $D\geq 1$, 
the free energy is not defined, since Eq. (112) diverges. This 
shows that perturbation theory with respect to the coupling 
parameter, generally speaking, fails for the disordered system.

In order to develop a perturbation theory, some special 
tricks are necessary. For example, the disorder strength and 
the interaction strength could be treated not as independent 
quantities, but as being related through the ratio
\be
\label{116}
\lbd \equiv \frac{\xi_0^2}{2g} \; ,
\ee
which is assumed to be fixed [60]. Then the free energy (112), 
for the Gaussian distribution (109), takes the form
\be
\label{117}
f(g,\lbd) = -\; \frac{1}{\sqrt{4\pi g\lbd}} \; 
\int_{-\infty}^{\infty} \;\exp\left (- \; 
\frac{\xi^2}{4g\lbd} \right ) \; \ln Z(g,\xi) \; d\xi \; .
\ee
This expression can be expanded in powers of $g$, yielding in 
the $k$-order
\be
\label{118}
f_k(g,\lbd) = \sum_{n=1}^k a_n(\lbd) g^n \; .
\ee
The coefficients $a_n(\lbd)$ are not easy to define, and for 
$0\leq\lbd\leq 1$ they were found to be represented as the sum
\be
\label{119}
a_n(\lbd) = b_n(\lbd) + c_n(\lbd) \; ,
\ee
whose terms for $n\gg 1$ are [60]
$$
b_n(\lbd) = (-1)^{n+1} \; 
\frac{4^n(n-1)!}{\sqrt{2}\;\pi} \; (1 -\lbd)^{n-1/2} \; ,
$$
$$
c_n(\lbd) = \frac{4^n n!\lbd^n}{\sqrt{\pi}\; n^{3/2}} \; 
\exp \left (-\gm\sqrt{n} +\al\right ) 
\cos(\left ( \mu\sqrt{n} +\bt\right ) \; ,
$$
where
$$
\gm = \frac{2.959237}{\sqrt{\lbd}} \; , \qquad
\mu = \frac{3.184867}{\sqrt{\lbd}} \; , \qquad
\al = \frac{\ln 2}{4\lbd} \; , \qquad 
\bt = \frac{3\pi}{4\lbd} \; .
$$
Although the found coefficients $a_n=a_n(\lbd)$ are valid only 
for $n\gg 1$, formally using them for all $n\geq 1$, and fixing 
$\lbd=0.1$, gives
$$
a_1 =0.854049\; , \qquad a_2 = -3.07481 \; , \qquad a_3=22.1387 \; ,
$$
$$
a_4 = -239.098 \; , \qquad a_5 = 3443 \; , \qquad a_6=-61974.1 \; ,
$$
and so on. These coefficients grow very fast, making series (118) 
strongly divergent. The series were shown [60] to be so widely 
divergent that they could not be Borel summed.

But the divergent series (118) can be summed employing the 
self-similar approximation theory [48--50]. We have accomplished 
the resummation procedure by using the self-similar factor 
approximants [51--54]. The latter have the form
\be
\label{120}
f_k^*(g,\lbd) = a_1(\lbd) g \prod_{i=1}^{N_k}  ( 1 + A_i g)^{n_i} \; ,
\ee
where
\begin{eqnarray}
\nonumber
N_k = \left \{ \begin{array}{ll}
k/2 , & \; k=2,4,\ldots \\
(k+1)/2, & \; k=3,5,\ldots 
\end{array} \right.
\end{eqnarray}
The coefficients $A_i$ and $n_i$ are defined by re-expanding form 
(120) and equating it to series (118), which yields
$$
\sum_{i=1}^{N_k} n_i A_i^n = B_n \qquad (n=1,2,\ldots, k) \; ,
$$
$$
B_n = \frac{(-1)^{n-1}}{(n-1)!} \; \lim_{g\ra 0} \;
\frac{d^n}{d g^n} \; \ln \; \left [
\frac{f_k(g,\lbd)}{a_1(\lbd)g} \right ] \; ,
$$
and $A_1=1$ for the odd $N_k=(k+1)/2$. For the fixed $\lbd=0.1$ 
and different $g$, we calculated the factor approximants (120) 
up to the order $k=10$. For $g=0.1$, we get $f_{10}^*=0.0662$, 
with the relative error $22\%$ as compared to the exact value 
$0.0542$. For $g=1$, we have $f_{10}^*=0.229$, with an error 
$19\%$, as compared to the exact value $0.251$. And for $g=10$, 
we find $f_{10}^*=0.712$, deviating from the exact value $0.642$ 
by $11\%$. These approximants should be considered as quite good, 
if one remembers that expressions (119) have been defined only 
for $n\gg 1$, but used for all $n=0,1,2,\ldots,10$. Therefore 
the resulting errors rather characterize the inaccuracy of these 
coefficients (119), but not merely the accuracy of the factor 
approximants (120).

The main message of this Section is that, under disorder, 
the perturbation theory with respect to atomic interactions 
can become inapplicable. For the Gaussian distribution of 
the disorder potential, this perturbation theory fails for any 
strength of disorder, and for the uniform disorder distribution, 
the perturbation theory becomes invalid at sufficiently strong 
disorder. To realize the weak-coupling perturbation theory, it 
is necessary to invoke some tricks, like fixing a relation between 
the disorder strength and the interaction strength. However, 
resorting to such tricks looks too artificial, since in reality 
the disorder and interaction strength are independent 
characteristics.

\section{Creation of Nonequilibrium Granular Condensate}

In the previous Sections, we have considered equilibrium 
systems with Bose-Einstein condensate, which, by applying 
spatially random fields, could be transformed into a Bose-glass 
type phase with a granular condensate. In the present Section, 
we show that a Bose system, subject to the action of an external 
time-alternating potential enjoys the properties to some extent 
analogous to those of a system in a spatially random potential. 
In particular, in such a nonequilibrium system a nonequilibrium 
granular condensate can be created.

Let us consider a Hamiltonian
\be
\label{121}
H(t) = \hat H + \hat V(t) \; ,
\ee
in which the first term
\be
\label{122}
\hat H = \int \psi^\dgr(\br) \left [ -\; 
\frac{\nabla^2}{2m} + U(\br) \right ] \; \psi(\br)\; d\br \; 
+ \; \frac{\Phi_0}{2} 
\int \psi^\dgr(\br) \psi^\dgr(\br) \psi(\br)\psi(\br) \; d\br 
\ee
describes Bose atoms in a trapping potential $U(\br)$, and the 
second term
\be
\label{123}
\hat V(t) = \int \psi^\dgr(\br) V(\br,t) \psi(\br) \; d\br 
\ee
corresponds to an external alternating potential. Again, for 
brevity we do not write explicitly the time dependence of the 
field operators $\psi(\br)=\psi(\br,t)$.

The external potential $V(\br,t)$ is assumed to vary in time so 
that its characteristic variation time $t_{var}$ is much larger 
than the local-equilibrium time $t_{loc}$, but much shorter than 
the observation time,
\be
\label{124}
t_{loc} \ll t_{var} \ll t_{obs} \; .
\ee
As an estimate of the local-equilibrium time, we can take the 
values of the parameters typical of the dilute trapped gases 
$^{87}$Rb and $^{23}$Na, say, $m\sim 10^{-22}$ g, $\rho\sim 10^{15}$ 
cm$^{-3}$, and $a_s\sim 10^{-7}$ cm. The local-equilibrium time is 
defined [3] as
\be
\label{125}
t_{loc} = \frac{m}{\rho a_s} \; .
\ee
With the above parameters, this gives $t_{loc}\sim 10^{-3}$ s.

Under condition (124), it is possible to define the local-equilibrium 
free energy
\be
\label{126}
F(t) = - T(t) \ln\; {\rm Tr} e^{-\bt(t) H(t)} \; ,
\ee
in which $T(t)\equiv1/\bt(t)$ is the local-equilibrium temperature. 
Since the observation time is much longer than the local-equilibrium 
time (125), it is appropriate to introduce the average free energy
\be
\label{127}
F = \frac{1}{t_{var}} \; \int_0^{t_{var}} \; F(t)\; dt \; ,
\ee
averaging the local-equilibrium free energy (126) over the 
characteristic variation time of the alternating potential $V(\br,t)$.

Suppose the alternating potential has the form
\be
\label{128}
V(\br,t) = V(\br) f(t) \; ,
\ee
in which the temporal function $f(t)$ varies in the interval
\be
\label{129}
f_{min} \leq f(t) \leq f_{max} \; .
\ee
Let us define the function $t=t(\xi)$ by the equation
\be
\label{130}
f(t(\xi)) = \xi \; .
\ee
By introducing the effective Hamiltonian
\be
\label{131}
H_{eff}(\xi) \equiv H(t(\xi)) \; ,
\ee
we get
\be
\label{132}
H_{eff}(\xi) = \hat H + \int \psi(\br) \xi V(\br) \psi(\br) \; d\br \; .
\ee
Also, we define the effective temperature
\be
\label{133}
T_{eff}(\xi) \equiv T(t(\xi)) \equiv \frac{1}{\bt_{eff}(\xi)} \; .
\ee
The quantity
\be
\label{134}
p(\xi) \equiv \frac{1}{t_{var}} \; \left | \frac{dt(\xi)}{d\xi} 
\right | 
\ee
plays the role of the distribution of the variable $\xi$. With these 
notations, the free energy (127) transforms into
\be
\label{135}
F = - \int_{f_{min}}^{f_{max}} \; T_{eff}(\xi) \;
\ln \; {\rm Tr}\; \exp \left \{ -\bt_{eff}(\xi) H_{eff}(\xi) 
\right \} \; p(\xi) \; d\xi \; .
\ee
Let $\xi^*$ be a value in the interval $f_{min}\leq\xi^*\leq f_{max}$,
and let us define
\be
\label{136}
T^* \equiv T_{eff}(\xi^*) \equiv \frac{1}{\bt^*} \; .
\ee
Then, by the theorem of mean, the free energy (135) can be approximately 
represented as
\be
\label{137}
F \cong -T^* \int_{f_{min}}^{f_{max}} 
\ln \; {\rm Tr}\; \exp \left \{ -\bt^* H_{eff}(\xi) 
\right \} \; p(\xi) \; d\xi \; .
\ee
This expression looks analogously to the free energy of a system in an 
external random potential $\xi V(\br)$ that enters Eq. (132).

To exemplify explicitly the form of distribution (134), let us take the 
time-dependent factor of the alternating potential (128) as
\be
\label{138}
f(t) = \cos(\om t) \; .
\ee
The characteristic variation time here is, clearly, the period 
$t_{var}=2\pi/\om$. The variation range of function (138), as defined 
in Eq. (129), is given by $f_{min}=-1$ and $f_{max}=1$. The function 
$t(\xi)$ is found from Eq. (130), yielding
\be
\label{139}
t(\xi) = \frac{1}{\om} \; {\rm arc cos} \xi \; .
\ee
>From here, for distribution (134), we get
\be
\label{140}
p(\xi) = \frac{1}{\pi\sqrt{1-\xi^2}} \qquad ( - 1 \leq \xi \leq 1) \; .
\ee
The variable $\xi$, with distribution (140), is zero-centered, since
$$
\int_{-1}^{1} \xi p(\xi)\; d\xi = 0 \; ,
$$
and its dispersion is
$$
\Dlt^2(\xi) = \int_{-1}^{1} \xi^2 p(\xi)\; d\xi = \frac{1}{2} \; .
$$

In this way, when considering the time-averaged behavior of a Bose 
system, subject to the action of an external alternating potential, we 
come to the picture that is similar to the description of a Bose system 
in an external spatially random potential. The consequences, therefore, 
should also be similar. The overall behavior of trapped atoms, in the 
presence of an alternating potential $V(\br,t)$, should be as follows.

Let $V_0$ be the strength of the alternating potential $V(\br,t)
\sim V_0$. The latter plays the role of $V_R$ in the localization length 
(84). If the whole trap is perturbed by the alternating potential, then 
the characteristic trap length $l_0=1/\sqrt{m\om_0}$, where $\om_0$ is 
the trap frequency, plays the role of $l_R$ in Eq. (84). Hence, for the
localization length (84), we have
$$
l_{loc} \sim \left ( \frac{\om_0}{V_0} \right )^2 l_0 \; .
$$
When the amplitude of the alternating potential is small, such that 
$V_0\ll\om_0$, then there is an extended condensate filling the trap. 
If the alternating potential oscillates with a frequency far detuned 
from any transition frequency for trapped atoms, then the extended 
condensate is just a perturbed ground-state condensate. But if the 
frequency of the oscillating potential is in resonance with the 
transition frequency between the ground-state condensate and an 
excited topological mode, then the extended condensate is formed by 
the fluctuating ground-state condensate mode and an excited coherent 
mode [61--64].

Increasing the amplitude $V_0$ of the alternating potential perturbs 
the Bose-condensed system stronger. When $V_0$ reaches the trap frequency 
$\om_0$, the localization length becomes comparable with the effective 
trap length $l_0$. At this moment, the granular condensate starts being 
formed. Since the applied external field is time dependent, the created 
granular condensate is nonequilibrium. This means that the disconnected 
regions with the condensate, with time, change their shapes, disappear 
and appear again, so that on average there always exist several such 
regions of the granular condensate.

The nonequilibrium granular condensate can exist for the amplitude of 
the alternating potential in the range
$$
\om_0 \leq V_0 \leq \om_0\; \sqrt{ \frac{l_0}{a} } \; .
$$
If the amplitude is so large that $V_0\sim\om_0\sqrt{l_0/a}$, then 
$l_{loc}\sim a$, and the granular condensate is completely destroyed. 
Then the whole system is a strongly nonequilibrium normal matter, 
containing no condensate, but being rather in a normal turbulent state.

The phase portrait of the nonequilibrium system, at zero temperature, 
qualitatively corresponds to that of the equilibrium random system, as 
is shown in Fig. 1, with the following analogies between the systems:
\vskip 2mm

{\it equilibrium extended condensate $\leftrightarrow$ nonequilibrium 
extended condensate, 
\vskip 2mm
equilibrium granular condensate $\leftrightarrow$ 
nonequilibrium granular condensate,
\vskip 2mm
equilibrium normal glass $\leftrightarrow$ 
turbulent normal fluid.
}

\vskip 2mm

The principal difference between the equilibrium and nonequilibrium 
granular condensates is as follows. The {\it equilibrium} granular 
condensate represents a system with randomly distributed in space 
regions of the condensate, surrounded by the normal phase with no 
condensate, these regions being stationary in time, their location 
in space being fixed. The {\it nonequilibrium} granular condensate, 
at each instant of time, is analogous to its equilibrium counterpart.
However, the regions with the condensate are not fixed in time. But, 
as time varies, the condensate regions change their shapes and 
locations. They can appear in new spatial locations, but disappear 
in others. What is fixed for a nonequilibrium granular condensate is 
the average (over space and time) concentration of condensed atoms, 
so that the average condensate fraction $n_0$ remains a well defined 
order parameter. At each instant of time, the system with the 
nonequilibrium granular condensate is a kind of a snapshot, which 
is analogous to the equilibrium granular condensate. Therefore, if 
trapped atoms are released from the trap, the time-of-flight observations 
for a nonequilibrium system will be similar to those for an equilibrium 
system.

We may note that the nature of bosons, considered above, can be any. 
These could be usual bosonic atoms [2--5]. Or these could be composite 
bosonic molecules formed of fermions [3,65].

The discussed analogies between the random Bose-condensed systems 
and the nonequlibrium Bose systems can be tested experimentally. 
Such experiments are now in progress in the Institute of Physics of 
S\~ao Carlos, University of S\~ao Paulo, Brazil. Preliminary results 
confirm these analogies. But a detailed exposition of experiments will 
be done in separate publications.

\vskip 5mm

{\bf Acknowledgement}

\vskip 2mm
Financial support from the Russian Foundation for Basic Research 
(Grant 08-02-00118) is acknowledged.

\newpage

\begin{figure}[ht]
\centerline{\psfig{file=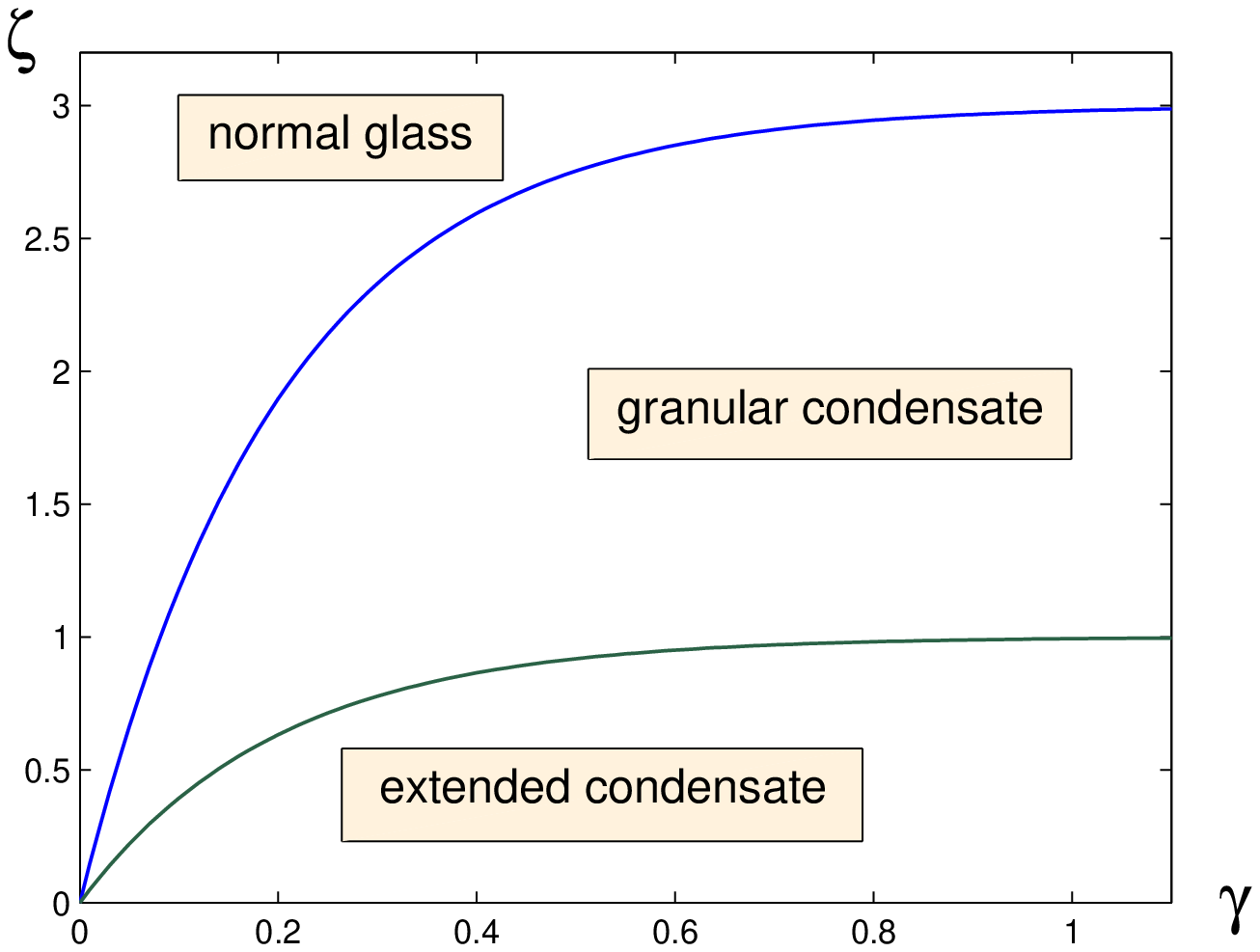,height=5in}}
\caption{Qualitative phase portrait of a Bose system in a spatially 
random potential at zero temperature on the plane of the gas parameter 
$\gm$ and disorder parameter $\zeta$.}
\label{fig:Fig.1}
\end{figure}

\end{document}